\documentclass[a4paper,twoside,reqno]{bjp}
\usepackage{graphicx}
\usepackage{cite}
\usepackage{amssymb,amsmath,amscd,amsthm}
\usepackage{times}
\usepackage{physics}

\usepackage[bookmarks=false]{hyperref}
\hypersetup{%
    colorlinks=true,        
    linkcolor=blue,          
    citecolor=blue,         
    urlcolor=blue           
    }

\usepackage{geometry}
\allowdisplaybreaks

\begin{document}

\title{Connection
between Elliott SU(3) and spherical shell model bases}

\runningheads{A. Martinou et al.}{Elliott SU(3) and the shell model}

\begin{start}{%
\author{A. Martinou}{1},
\author{N. Minkov}{2}
\author{S. Sarantopoulou}{1}
\author{S. Peroulis}{1}
\author{I. E. Assimakis}{1}
\author{D. Bonatsos}{1}

\address{Institute of Nuclear and Particle Physics, National Centre for Scientific Research ``Demokritos'', GR-15310 Aghia Paraskevi, Attiki, Greece.}{1}
\address{ Institute of Nuclear Research and Nuclear Energy, Bulgarian Academy of Sciences, 72 Tzarigrad Road, 1784 Sofia, Bulgaria.}{2}

\received{Day Month Year (Insert date of submission)}
}

\begin{Abstract}
In the Elliott SU(3) symmetry scheme the single particle basis is derived from the isotropic harmonic oscillator Hamiltonian in the Cartesian coordinate system. These states are transformed into the solutions of the same Hamiltonian within the spherical coordinate system. Then the spin-orbit coupling can be added in a straightforward way. The outcome is a transformation between the Elliott single particle basis and the spherical shell model space. 
\end{Abstract}


\end{start}


\section{Introduction}
Nuclear shape is the important factor which determines the appropriate good quantum numbers of the single particle states and furthermore all the collective nuclear observables. Usually the nucleus is considered to possess a spherical  or an ellipsoidal (axial or triaxial) shape. If the nucleus is nearly spherical, then the spherical coordinate system $(r,\theta,\phi)$ is the appropriate one for the mathematical solution of the Schr\"odinder equation. In contrast, if the nucleus is triaxial, then the Cartesian coordinate system $(x,y,z)$ suits better to the single particle states. The nucleons are bound due to a mean field, which can be considered to be a three-dimensional (3D) isotropic harmonic oscillator. In addition there are other important interactions, such as the spin-orbit ${\bf l}\cdot {\bf s}$ \cite{Mayer1,Mayer2}, the quadrupole-quadrupole ${\bf q}\cdot {\bf q}$, and the pairing interaction. The spin-orbit and the pairing interactions use the angular momentum $l$, thus they match better  to the spherical coordinate system. On the contrary the five components of the ${\bf q}\cdot {\bf q}$ interaction are described more easily in the Cartesian coordinate system \cite{Harvey}. Consequently one has to interplay between the two coordinate systems.

\section{The bases}

Let us consider the simplest single particle Hamiltonian:
\begin{equation}
H_i={p_i^2\over 2m}+{1\over 2}m\omega^2r_i^2+V_{ls} {\bf l} \cdot {\bf s},
\end{equation}
where the first two terms are the Hamiltonian of the 3D isotropic harmonic oscillator and the last term is the spin-orbit interaction, with ${\bf l}$ being the orbital angular momentum and ${\bf s}$ being the spin, while $V_{ls}$ is the potential of the spin-orbit coupling \cite{Greiner}. 

The single particle states of the Elliott SU(3) scheme \cite{Elliott1,Elliott2,Elliott3} are the solutions of the 3D isotropic harmonic oscillator in the Cartesian coordinate system $\ket{n_z,n_x,n_y}$, which are Hermite polynomials \cite{Cohen}. These states have been used in Ref. \cite{Draayer89} to derive the U(3) irreducible representations (irreps) $[f_1,f_2,f_3]$ and afterwards the Elliott irreps $(\lambda, \mu)$.

The 3D isotropic harmonic oscillator can also be solved in the spherical coordinate system and give single particle states $\{k,l,l_0\}$ \cite{Cohen}, where the total number of oscillator quanta $\mathcal{N}$ is:
\begin{equation}
\mathcal{N}=k+l,
\end{equation}
with $l_0$ being the projection of the orbital angular momentum, while $k$ is an even positive integer or zero. In nuclear physics the notation $k=2n$ is also used, with $n=0$, 1, 2, \dots, resulting in $\mathcal{N}=2n+l$.

\section{Transformation of the 3D isotropic harmonic oscillator states}\label{trans}

A 3D harmonic oscillator shell with $\mathcal{N}=1$ has single particle states in the Cartesian coordinate system $\ket{n_z,n_x,n_y}$:
\begin{equation}
\ket{n_z,n_x,n_y}: \ket{1,0,0},\ket{0,1,0},\ket{0,0,1}.
\end{equation}

For the $p$ shell the possible $\{k,l,l_0\}$ are:
\begin{equation}
\{k,l,l_0\}:\{0,1,0\},\{0,1,1\},\{0,1,-1\}.
\end{equation}
The relations between the Cartesian and the spherical coordinate systems for $\mathcal{N}=1$ are \cite{Cohen}:
\begin{eqnarray}
\ket{1,0,0}=\{0,1,0\},\label{t1}\\
\ket{0,1,0}={1\over \sqrt{2}}(\{0,1,-1\})-\{0,1,1\}),\label{t2}\\
\ket{0,0,1}={i\over\sqrt{2}}(\{0,1,-1\})+\{0,1,1\}).\label{t3}
\end{eqnarray}

Similar transformations are given in \cite{Cohen} for the 3D isotropic harmonic oscillator states with $\mathcal{N}=2$:
\begin{eqnarray}
\ket{2,0,0}=-{1\over\sqrt{3}}\{2,0,0\}+\sqrt{2\over3}\{0,2,0\},\label{c1}\\
\ket{1,1,0}=-{1\over\sqrt{2}}\{0,2,1\}+{1\over\sqrt{2}}\{0,2,-1\},\label{c2}\\
\ket{1,0,1}={i\over\sqrt{2}}\{0,2,1\}+{i\over\sqrt{2}}\{0,2,-1\},\label{c3}\\
\ket{0,2,0}=-{1\over\sqrt{3}}\{2,0,0\}+{1\over 2}\{0,2,2\}+ 
{1\over 2}\{0,2,-2\}-{1\over\sqrt{6}}\{0,2,0\},\label{c4}\\
\ket{0,1,1}=-{i\over\sqrt{2}}\{0,2,2\}+{i\over\sqrt{2}}\{0,2,-2\},\label{c5}\\
\ket{0,0,2}=-{1\over\sqrt{3}}\{2,0,0\}-{1\over 2}\{0,2,2\}-
{1\over 2}\{0,2,-2\}-{1\over \sqrt{6}}\{0,2,0\}.\label{c6}
\end{eqnarray}

\section{Clebsch-Gordan coefficients}\label{CG}

In the following the ${\bf j}={\bf l}+{\bf s}$ coupling is applied in each $\{k,l,l_0\}$ state, with ${\bf j}$ being the total angular momentum. The notation $\ket{s,s_0}=\ket{1/2,\pm 1/2}=\ket{\pm}$ is used, where $s_0$ is the spin projection. The Clebsch-Gordan (CG) coefficients connect the uncoupled states $\{k,l,l_0\}\ket{\pm}$ with the coupled states $\{k,l,j,j_0\}$ (where $j_0$ is the projection of the total angular momentum), as:
\begin{equation}
\{k,l,l_0\}\ket{\pm}=\sum_j\sum_{j_0}C_{l,l_0,\pm}^{j,j_0}\{k,l,j,j_0\}. 
\end{equation}

The coupled states $\{k,l,j,j_0\}$ are the usual shell model orbitals:
\begin{eqnarray}
\{0,1,1/2,\pm1/2\}=1p^{j=1/2}_{j_0=\pm 1/2},\\
\{0,1,3/2,\pm 1/2\}=1p^{j=3/2}_{j_0=\pm 1/2},\\
\{0,1,3/2,\pm 3/2\}=1p^{j=3/2}_{j_0=\pm 3/2},\\
\{2,0,1/2,\pm 1/2\}=2s^{j=1/2}_{j_0=\pm 1/2},\\
\{0,2,j,j_0\}=1d^{j}_{j_0}.
\end{eqnarray}

Calculating the CG coefficients ones finds:
\begin{eqnarray}
\{0,1,0\}\ket{+}=-{1\over \sqrt{3}}1p^{1/2}_{1/2}+\sqrt{2\over 3}1p^{3/2}_{1/2},\label{s1}\\
\{0,1,0\}\ket{-}={1\over \sqrt{3}}1p^{1/2}_{-1/2}+\sqrt{2\over 3}1p^{3/2}_{-1/2},\label{s2}\\
\{0,1,1\}\ket{+}=1p^{3/2}_{3/2},\label{s3}\\
\{0,1,1\}\ket{-}=\sqrt{2\over 3}1p^{1/2}_{1/2}+{1\over \sqrt{3}}1p^{3/2}_{1/2},\label{s4}\\
\{0,1,-1\}\ket{+}=-\sqrt{2\over 3}1p^{1/2}_{-1/2}+{1\over \sqrt{3}}1p^{3/2}_{-1/2},\label{s5}\\
\{0,1,-1\}\ket{-}=1p^{3/2}_{-3/2}.\label{s6}
\end{eqnarray}

The same connection can be achieved for the $sd$ shell with $\mathcal{N}=2$:
\begin{eqnarray}
\{2,0,0\}\ket{+}=2s^{1/2}_{1/2},\label{sp1}\\
\{2,0,0\}\ket{-}=2s^{1/2}_{-1/2},\label{sp2}\\
\{0,2,2\}\ket{+}=1d^{5/2}_{5/2},\label{sp3}\\
\{0,2,2\}\ket{-}={2\over\sqrt{5}}1d^{3/2}_{3/2}+{1\over\sqrt{5}}1d^{5/2}_{3/2},\label{sp4}\\
\{0,2,-2\}\ket{+}=-{2\over\sqrt{5}}1d^{3/2}_{-3/2}+{1\over\sqrt{5}}1d^{5/2}_{-3/2},\label{sp5}\\
\{0,2,-2\}\ket{-}=1d^{5/2}_{-5/2},\label{sp6}\\
\{0,2,1\}\ket{+}=-{1\over\sqrt{5}}1d^{3/2}_{3/2}+{2\over\sqrt{5}}1d^{5/2}_{3/2},\label{sp7}\\
\{0,2,1\}\ket{-}=\sqrt{3\over 5}1d^{3/2}_{1/2}+\sqrt{2\over 5}1d^{5/2}_{1/2},\label{sp8}\\
\{0,2,-1\}\ket{+}=-\sqrt{3\over 5}1d^{3/2}_{-1/2}+\sqrt{2\over 5}1d^{5/2}_{-1/2},\label{sp9}\\
\{0,2,-1\}\ket{-}={1\over\sqrt{5}}1d^{3/2}_{-3/2}+{2\over\sqrt{5}}1d^{5/2}_{-3/2},\label{sp10}\\
\{0,2,0\}\ket{+}=-\sqrt{2\over 5}1d^{3/2}_{1/2}+\sqrt{3\over 5}1d^{5/2}_{1/2},\label{sp11}\\
\{0,2,0\}\ket{-}=\sqrt{2\over 5}1d^{3/2}_{-1/2}+\sqrt{3\over 5}1d^{5/2}_{-1/2}.\label{sp12}
\end{eqnarray}

\section{Transformation between the Elliott and the shell model bases}\label{Shell}
 
The combination of Eqs. (\ref{t1}-\ref{t3}) with (\ref{s1}-\ref{s6}) connects the Elliott single particle states $\ket{n_z,n_x,n_y}\ket{\pm}$ with the shell model states for the $p$ shell ($\mathcal{N}=1$) as:
\begin{eqnarray}
\ket{1,0,0}\ket{+}=-{1\over \sqrt{3}}1p^{1/2}_{1/2}+\sqrt{2\over 3}1p^{3/2}_{1/2},\nonumber\\\label{p1}
\\
\ket{1,0,0}\ket{-}={1\over \sqrt{3}}1p^{1/2}_{-1/2}+\sqrt{2\over 3}1p^{3/2}_{-1/2},\nonumber\\\\\label{p2}
\ket{0,1,0}\ket{+}=-{1\over \sqrt{3}}1p^{1/2}_{-1/2}+{1\over\sqrt{6}}1p^{3/2}_{-1/2}-{1\over\sqrt{2}}1p^{3/2}_{3/2},\nonumber\\\\\label{p3}
\ket{0,1,0}\ket{-}={1\over\sqrt{2}}1p^{3/2}_{-3/2}-{1\over\sqrt{3}}1p^{1/2}_{1/2}-{1\over\sqrt{6}}1p^{3/2}_{1/2},\nonumber\\\\\label{p4}
\ket{0,0,1}\ket{+}=-{i\over\sqrt{3}}1p^{1/2}_{-1/2}+{i\over\sqrt{6}}1p^{3/2}_{-1/2}+{i\over\sqrt{2}}1p^{3/2}_{3/2},\nonumber\\\\\label{p5}
\ket{0,0,1}\ket{-}={i\over\sqrt{2}}1p^{3/2}_{-3/2}+{i\over\sqrt{3}}1p^{1/2}_{1/2}+{i\over\sqrt{6}}1p^{3/2}_{1/2}.\nonumber\\\label{p6}
\end{eqnarray}

The transformation between the Elliott and the shell model bases can also be established in the $sd$ shell with $\mathcal{N}=2$ quanta. The system of Eqs. (\ref{c1}-\ref{c6}) with the Eqs. (\ref{sp1}-\ref{sp12}) results to:
\begin{eqnarray}
\ket{2,0,0}\ket{+}=-{1\over\sqrt{3}}2s^{1/2}_{1/2}-{2\over\sqrt{15}}1d^{3/2}_{1/2}+\sqrt{2\over 5}1d^{5/2}_{1/2},\nonumber\\\label{sd1}\\
\ket{2,0,0}\ket{-}=-{1\over\sqrt{3}}2s^{1/2}_{-1/2}+{2\over\sqrt{15}}1d^{3/2}_{-1/2}+\sqrt{2\over 5}1d^{5/2}_{-1/2},\nonumber\\\\\label{sd2}
\ket{1,1,0}\ket{+}={1\over\sqrt{10}}1d^{3/2}_{3/2}-\sqrt{3\over 10}1d^{3/2}_{-1/2}-\sqrt{2\over 5}1d^{5/2}_{3/2} 
+{1\over\sqrt{5}}1d^{5/2}_{-1/2},\nonumber\\\\\label{sd3}
\ket{1,1,0}\ket{-}=-\sqrt{3\over 10}1d^{3/2}_{1/2}+{1\over\sqrt{10}}1d^{3/2}_{-3/2} 
-{1\over\sqrt{5}}1d^{5/2}_{1/2}+\sqrt{2\over 5}1d^{5/2}_{-3/2},\nonumber\\\\\label{sd4}
\ket{1,0,1}\ket{+}=-{i\over\sqrt{10}}1d^{3/2}_{3/2}-i\sqrt{3\over 10}1d^{3/2}_{-1/2} 
+i\sqrt{2\over 5}1d^{5/2}_{3/2}+{i\over\sqrt{5}}1d^{5/2}_{-1/2},\nonumber\\\\\label{sd5}
\ket{1,0,1}\ket{-}=i\sqrt{3\over 10}1d^{3/2}_{1/2}+{i\over\sqrt{10}}1d^{3/2}_{-3/2} 
+{i\over\sqrt{5}}1d^{5/2}_{1/2}+i\sqrt{2\over 5}1d^{5/2}_{-3/2},\nonumber\\\\\label{sd6}
\ket{0,2,0}\ket{+}=-{1\over\sqrt{3}}2s^{1/2}_{1/2}+{1\over\sqrt{15}}1d^{3/2}_{1/2}-{1\over\sqrt{5}}1d^{3/2}_{-3/2}\nonumber\\+{1\over 2}1d^{5/2}_{5/2}-{1\over\sqrt{10}}1d^{5/2}_{1/2}+{1\over\sqrt{20}}1d^{5/2}_{-3/2}, \nonumber\\\\\label{sd7}
\ket{0,2,0}\ket{-}=-{1\over\sqrt{3}}2s^{1/2}_{-1/2}+{1\over\sqrt{5}}1d^{3/2}_{3/2}-{1\over\sqrt{15}}1d^{3/2}_{-1/2}\nonumber\\+{1\over\sqrt{20}}1d^{5/2}_{3/2}
-{1\over\sqrt{10}}1d^{5/2}_{-1/2}+{1\over 2}1d^{5/2}_{-5/2},\nonumber\\\\\label{sd8}
\ket{0,1,1}\ket{+}=-i\sqrt{2\over 5}1d^{3/2}_{-3/2}-{i\over\sqrt{2}}1d^{5/2}_{5/2} 
+{i\over\sqrt{10}}1d^{5/2}_{-3/2},\nonumber\\\\\label{sd9}
\ket{0,1,1}\ket{-}=-i\sqrt{2\over 5}1d^{3/2}_{3/2}-{i\over\sqrt{10}}1d^{5/2}_{3/2}
+{i\over\sqrt{2}}1d^{5/2}_{-5/2},\nonumber\\\\\label{sd10}
\ket{0,0,2}\ket{+}=-{1\over\sqrt{3}}2s^{1/2}_{1/2}+{1\over\sqrt{15}}1d^{3/2}_{1/2}+{1\over\sqrt{5}}1d^{3/2}_{-3/2}\nonumber\\-{1\over 2}1d^{5/2}_{5/2}
-{1\over\sqrt{10}}1d^{5/2}_{1/2}-{1\over\sqrt{20}}1d^{5/2}_{-3/2},\nonumber\\\\\label{sd11}
\ket{0,0,2}\ket{-}=-{1\over\sqrt{3}}2s^{1/2}_{-1/2}-{1\over\sqrt{5}}1d^{3/2}_{3/2}-{1\over\sqrt{15}}1d^{3/2}_{-1/2}\nonumber\\-{1\over\sqrt{20}}1d^{5/2}_{3/2}
-{1\over\sqrt{10}}1d^{5/2}_{-1/2}-{1\over 2}1d^{5/2}_{-5/2}.\nonumber\\\label{sd12}
\end{eqnarray}

Eqs. (\ref{p1}-\ref{p6}) can also be solved in the inverse way. The shell model states are related with the Elliott states for the $\mathcal{N}=1$ shell as:
\begin{eqnarray}
1p^{1/2}_{1/2}=-{1\over\sqrt{3}}\ket{1,0,0}\ket{+}-{1\over\sqrt{3}}\ket{0,1,0}\ket{-} 
-{i\over\sqrt{3}}\ket{0,0,1}\ket{-},\\\label{pb1}
1p^{1/2}_{-1/2}={1\over\sqrt{3}}\ket{1,0,0}\ket{-}-{1\over\sqrt{3}}\ket{0,1,0}\ket{+} 
+{i\over\sqrt{3}}\ket{0,0,1}\ket{+},\\\label{pb2}
1p^{3/2}_{3/2}=-{1\over\sqrt{2}}\ket{0,1,0}\ket{+}-{i\over\sqrt{2}}\ket{0,0,1}\ket{+},\\\label{pb3}
1p^{3/2}_{1/2}=\sqrt{2\over 3}\ket{1,0,0}\ket{+}-{1\over\sqrt{6}}\ket{0,1,0}\ket{-} 
-{i\over\sqrt{6}}\ket{0,0,1}\ket{-},\\\label{pb4}
1p^{3/2}_{-1/2}=\sqrt{2\over 3}\ket{1,0,0}\ket{-}+{1\over\sqrt{6}}\ket{0,1,0}\ket{+} 
-{i\over\sqrt{6}}\ket{0,0,1}\ket{+},\\\label{pb5}
1p^{3/2}_{-3/2}={1\over\sqrt{2}}\ket{0,1,0}\ket{-}-{i\over\sqrt{2}}\ket{0,0,1}\ket{-}.\label{pb6}
\end{eqnarray}

The shell model states in relation to the Elliott states for the $sd$ shell with $\mathcal{N}=2$ are derived from the system of Eqs. (\ref{sd1}-\ref{sd12}):
\begin{eqnarray}
1s_{1/2}^{1/2}=-{1\over\sqrt{3}}\ket{2,0,0}\ket{+}-{1\over\sqrt{3}}\ket{0,2,0}\ket{+} 
-{1\over\sqrt{3}}\ket{0,0,2}\ket{+},\\\label{sdb1}
1s^{1/2}_{-1/2}=-{1\over\sqrt{3}}\ket{2,0,0}\ket{-}-{1\over\sqrt{3}}\ket{0,2,0}\ket{-} 
-{1\over\sqrt{3}}\ket{0,0,2}\ket{-},\\\label{sdb2}
1d^{3/2}_{3/2}={1\over\sqrt{10}}\ket{1,1,0}\ket{+}+{i\over\sqrt{10}}\ket{1,0,1}\ket{+}\nonumber\\+{1\over\sqrt{5}}\ket{0,2,0}\ket{-}+i\sqrt{2\over 5}\ket{0,1,1}\ket{-} 
-{1\over\sqrt{5}}\ket{0,0,2}\ket{-},\\\label{sdb3}
1d^{3/2}_{1/2}=-{2\over\sqrt{15}}\ket{2,0,0}\ket{+}-\sqrt{3\over 10}\ket{1,1,0}\ket{-}\nonumber\\-i\sqrt{3\over 10}\ket{1,0,1}\ket{-}+{1\over\sqrt{15}}\ket{0,2,0}\ket{+} 
+{1\over\sqrt{15}}\ket{0,0,2}\ket{+},\\\label{sdb4}
1d^{3/2}_{-1/2}={2\over\sqrt{15}}\ket{2,0,0}\ket{-}-\sqrt{3\over 10}\ket{1,1,0}\ket{+}\nonumber\\+i\sqrt{3\over 10}\ket{1,0,1}\ket{+}-{1\over\sqrt{15}}\ket{0,2,0}\ket{-} 
-{1\over\sqrt{15}}\ket{0,0,2}\ket{-},\\\label{sdb5}
1d^{3/2}_{-3/2}={1\over\sqrt{10}}\ket{1,1,0}\ket{-}-{i\over\sqrt{10}}\ket{1,0,1}\ket{-}\nonumber\\-{1\over\sqrt{5}}\ket{0,2,0}\ket{+}+i\sqrt{2\over 5}\ket{0,1,1}\ket{+} 
+{1\over\sqrt{5}}\ket{0,0,2}\ket{+},\\\label{sdb6}
1d^{5/2}_{5/2}={1\over 2}\ket{0,2,0}\ket{+}+{i\over \sqrt{2}}\ket{0,1,1}\ket{+} 
-{1\over 2}\ket{0,0,2}\ket{+},\\\label{sdb7}
1d^{5/2}_{3/2}=-\sqrt{2\over 5}\ket{1,1,0}\ket{+}-i\sqrt{2\over 5}\ket{1,0,1}\ket{+}\nonumber\\+{1\over\sqrt{20}}\ket{0,2,0}\ket{+}+{i\over\sqrt{10}}\ket{0,1,1}\ket{-} 
-{1\over\sqrt{20}}\ket{0,0,2}\ket{-},\\\label{sdb8}
1d^{5/2}_{1/2}=\sqrt{2\over 5}\ket{2,0,0}\ket{+}-{1\over\sqrt{5}}\ket{1,1,0}\ket{-}\nonumber\\-{i\over\sqrt{5}}\ket{1,0,1}\ket{-}-{1\over\sqrt{10}}\ket{0,2,0}\ket{+} 
-{1\over\sqrt{10}}\ket{0,0,2}\ket{+},\\\label{sdb9}
1d^{5/2}_{-1/2}=\sqrt{2\over 5}\ket{2,0,0}\ket{-}+{1\over\sqrt{5}}\ket{1,1,0}\ket{+}\nonumber\\-{i\over\sqrt{5}}\ket{1,0,1}\ket{+}-{1\over\sqrt{10}}\ket{0,2,0}\ket{-} 
-{1\over\sqrt{10}}\ket{0,0,2}\ket{-},\\\label{sdb10}
1d^{5/2}_{-3/2}=\sqrt{2\over 5}\ket{1,1,0}\ket{-}-i\sqrt{2\over 5}\ket{1,0,1}\ket{-}\nonumber\\+{1\over\sqrt{20}}\ket{0,2,0}\ket{+}-{i\over\sqrt{10}}\ket{0,1,1}\ket{+} 
-{1\over\sqrt{20}}\ket{0,0,2}\ket{+},\\\label{sdb11}
1d^{5/2}_{-5/2}={1\over 2}\ket{0,2,0}\ket{-}-{i\over\sqrt{2}}\ket{0,1,1}\ket{-} 
-{1\over 2}\ket{0,0,2}\ket{-}.\label{sdb12}
\end{eqnarray}

Similar transformations for shells with $\mathcal{N}\ge 3$ are the subject of work in progress.

\section{Proxy-SU(3) in the shell model}
 
The proxy-SU(3) symmetry has been introduced in \cite{Proxy1}, using a replacement of the intruder Nilsson  orbitals in the spin-orbit like shells 28-50, 50-82, 82-126. Here we exemplify this replacement for the simpler case of the spin-orbit like shell 6-14 \cite{hnps, Sorlin}, consisting of the normal parity orbitals $1p^{1/2}_{\pm 1/2}$ and the intruder orbitals $1d^{5/2}_{\pm 5/2,\pm3/2,\pm 1/2}$. The SU(3) symmetry can be restored in this shell by excluding the highest $j_0$ orbitals $1d^{5/2}_{\pm 5/2}$ and by replacing the rest, $1d^{5/2}_{\pm 3/2,\pm 1/2}$, by the $1p^{3/2}_{\pm 3/2,\pm 1/2}$ respectively. This replacement simply reduces by one unit the number of quanta in the $z$ Cartesian axis of the intruder orbitals. Consequently the 6-14 shell, after the replacement of the $1d^{5/2}$ orbitals, becomes a $p$ shell, just like the 2-8 harmonic oscillator shell. In the following equations the orbitals of the 6-14 shell after this replacement are presented. The blue colored orbitals are those which were initially included in the original 6-14 shell before the replacement and the red colored orbitals are the orbitals included after the replacement. Thus proxy-SU(3) symmetry mixes the normal with the intruder parity orbitals for the 6-14 shell as follows:
\begin{eqnarray}
\ket{1,0,0}\ket{+}=-{1\over \sqrt{3}}1p^{1/2}_{1/2}+\sqrt{2\over 3}\underset{\color{blue}1d^{5/2}_{1/2}}{{\color{red}1p^{3/2}_{1/2}}},\nonumber\\\label{pc1}
\\
\ket{1,0,0}\ket{-}={1\over \sqrt{3}}1p^{1/2}_{-1/2}+\sqrt{2\over 3}\underset{\color{blue}1d^{5/2}_{-1/2}}{{\color{red}1p^{3/2}_{-1/2}}},\nonumber\\\\\label{pc2}
\ket{0,1,0}\ket{+}=-{1\over \sqrt{3}}1p^{1/2}_{-1/2}-{1\over\sqrt{2}}\underset{\color{blue}1d^{5/2}_{3/2}}{{\color{red}1p^{3/2}_{3/2}}}+{1\over\sqrt{6}}\underset{\color{blue}1d^{5/2}_{-1/2}}{{\color{red}1p^{3/2}_{-1/2}}},\nonumber\\\label{pc3}
\end{eqnarray}
\begin{eqnarray}
\ket{0,1,0}\ket{-}=-{1\over\sqrt{3}}1p^{1/2}_{1/2}-{1\over\sqrt{6}}\underset{\color{blue}1d^{5/2}_{1/2}}{{\color{red}1p^{3/2}_{1/2}}}+{1\over\sqrt{2}}\underset{\color{blue}1d^{5/2}_{-3/2}}{{\color{red}1p^{3/2}_{-3/2}}},\nonumber\\\\\label{pc4}
\ket{0,0,1}\ket{+}=-{i\over\sqrt{3}}1p^{1/2}_{-1/2}+{i\over\sqrt{2}}\underset{\color{blue}1d^{5/2}_{3/2}}{{\color{red}1p^{3/2}_{3/2}}}+{i\over\sqrt{6}}\underset{\color{blue}1d^{5/2}_{-1/2}}{{\color{red}1p^{3/2}_{-1/2}}},\nonumber\\\\\label{pc5}
\ket{0,0,1}\ket{-}={i\over\sqrt{3}}1p^{1/2}_{1/2}+{i\over\sqrt{6}}\underset{\color{blue}1d^{5/2}_{1/2}}{{\color{red}1p^{3/2}_{1/2}}}+{i\over\sqrt{2}}\underset{\color{blue}1d^{5/2}_{-3/2}}{{\color{red}1p^{3/2}_{-3/2}}}.\nonumber\\\label{pc6}
\end{eqnarray}

Eqs. (74)-(79) with the red terms are identical to Eqs. (38)-(43), which represent the Elliott $p$ shell, while Eqs. (74)-(79) with the blue terms correspond to the 6-14 spin-orbit like shell, which is mapped onto the $p$ shell through the proxy-SU(3) approximation.  

\section{An elementary example}

We describe  here  in the briefest possible way an elementary example of the usefulness of the equations derived in this work. More substantial examples will be presented in detail in a long publication.  
Consider the nucleus $^{20}_{10}$Ne$_{10}$. From the Nilsson diagrams one expects that the two valence protons and the two valence neutrons lie in the $1d^{5/2}_{\pm 1/2}$ orbitals. From the expression of these orbitals in terms of Elliott states, given in Eqs. (\ref{sdb9}), (\ref{sdb10}), one can calculate the average values of $\langle n_z\rangle$, $\langle n_x\rangle$, $\langle n_y\rangle$ for two particles in the $1d^{5/2}_{\pm 1/2}$ orbitals, which turn out to be 12/5, 4/5 and 4/5 respectively. These are average values over several states, thus it is not surprising that they are not integer. Then the {\sl effective average} Elliott quantum numbers for valence  protons or neutrons alone will be $\langle \lambda\rangle =\langle n_z\rangle -\langle n_x\rangle =8/5$ and $\langle \mu\rangle =\langle n_x\rangle -\langle n_y\rangle=0$ \cite{CRC}, while for the whole nucleus will be $\langle\lambda\rangle =16/5$ and $\langle \mu\rangle=0$. Again these are not integers, because they are average values over several states. Using the standard relations \cite{Draayer89} connecting $\lambda$ and $\mu$ to the collective parameters $\beta$ and $\gamma$  in the way employed in \cite{proxy2} we obtain $\beta=0.211$ and $\gamma=10.4^{\rm o}$. This parameter-independent result compares well with the relativistic mean field prediction of $\beta=0.186$ \cite{Lalazissis}.  

\section{Conclusions}

The Elliott basis is expressed in the Cartesian coordinate system, while the shell model space is described in the spherical coordinate system. A transformation has been established between the two bases for the $p$ and the $sd$ shells. For a deformed nucleus one may place the nucleons in the Elliott basis and use the above expansions in order to estimate the occupation probabilities of the shell model states. In contrast, for an almost spherical nucleus one may place the particles in the shell model states and use the expansions of the previous section, in order to estimate the collective deformation parameters $\beta$, $\gamma$ as in \cite{Castanos}, taking into account the probabilities by which  the various $ (\lambda, \mu)$ irreps appear.

\section*{Acknowledgements} 
Financial support by the Bulgarian National Science Fund (BNSF) under Contract No. KP-06-N28/6 is gratefully acknowledged.


\begin{thebibliography}{99}

\bibitem{Mayer1}
M. Goeppert-Mayer (1948) On Closed Shells in Nuclei \textit{Phys. Rev.} \textbf{74} 235-39 

\bibitem{Mayer2}
M. Goeppert-Mayer (1969) On Closed Shells in Nuclei II. \textit{Phys. Rev.} \textbf{75} 1969-70

\bibitem{Harvey}
M. Harvey (1968) The Nuclear SU(3) Model. In M. Baranger, and E. Vogt (eds) \textit{Advances in Nuclear Physics} vol \textbf{1}, New York: Plenum Press, pp. 67-180

\bibitem{Greiner}
W. Greiner, J. A. Maruhn (1996) \textit{Nuclear Models}, Springer-Verlag, Berlin-Heidelberg 

\bibitem{Elliott1}
J. P. Elliott  (1958) Collective motion in the nuclear shell model I. Classification schemes for states of mixed configurations \textit{Proc. Roy. Soc. Ser. A} \textbf{245} 128-45

\bibitem{Elliott2}
J. P. Elliott  (1958) Collective motion in the nuclear shell model II. The introduction of intrinsic wave-functions \textit{Proc. Roy. Soc. Ser. A} \textbf{245} 562-81

\bibitem{Elliott3}
J. P. Elliott, M. Harvey (1963). Collective motion in the nuclear shell model III. The calculation of spectra \textit{Proc. Roy. Soc. Ser. A} \textbf{272} 557-77

\bibitem{Cohen}
C. Cohen-Tannoudji, B. Diu, F. Laloe (1991) \textit{Quantum Mechanics} Vol. \textbf{1}. Wiley, Paris, Complement $B_{VII}$ pp. 814-23

\bibitem{Draayer89}
J. P. Draayer, Y. Leschber, S. C. Park, R. Lopez (1989) Representations of $U(3)$ in $U(N)$ \textit{Comp. Phys. Commun.} \textbf{56} 279-90


\bibitem{Proxy1} 
D. Bonatsos, I. E. Assimakis, N. Minkov et al. (2017) Proxy-SU(3) symmetry in heavy deformed nuclei \textit{Phys. Rev. C} \textbf{95} 064325

\bibitem{hnps}
A. Martinou, D. Bonatsos, N. Minkov et al. (2018) Nucleon numbers for nuclei with shape coexistence. \textit{Proceedings of the $27^{th}$ annual Symposium of the Hellenic Nuclear Physics Society} arXiv: 1810. 11860 [nucl-th]

\bibitem{Sorlin}
O. Sorlin, M.-G. Porquet (2008) Nuclear magic numbers: New features far from stability \textit{Prog. Part. Nucl. Phys.} \textbf{61} 602-73

\bibitem{CRC}
A. Martinou and D. Bonatsos (2019) Magic numbers of cylindrical symmetry arXiv: 1909.00233 [nucl-th]

\bibitem{Castanos}
O. Casta\~nos, J. P. Draayer and Y. Leschber (1988) Shape Variables and the Shell Model \textit{Z. Phys. A-Atomic Nuclei} \textbf{329} 33-43

\bibitem{proxy2}
D. Bonatsos, I. E. Assimakis, N. Minkov et al. (2017) Analytic predictions for nuclear shapes, the prolate dominance and the prolate-oblate shape transition in the proxy-SU(3) model \textit{Phys. Rev. C} \textbf{95} 064326

\bibitem{Lalazissis}
G. A. Lalazissis, S. Raman, and P. Ring (1999) Ground-state properties of even-even nuclei in the relativistic mean-field theory \textit{At. Data Nucl. Data Tables} {\textbf 71} 1  


\end{thebibliography}
\end{document}